\documentstyle[12pt,aaspp4]{article}
\def\ltsim{\, {}^<_\sim \,}
\def\gtsim{\, {}^>_\sim \,}
\def\etal{{\it et al.}}

\slugcomment{Submitted to Astron.J.}

\lefthead{G.~Harris \etal}
\righthead{The Halo of NGC 5128}

\begin{document}

\title{The Metallicity Distribution in the Halo Stars of \\
    NGC 5128:  Implications for Galaxy Formation}

\author{Gretchen L.~H.~Harris}
\affil{Department of Physics, University of Waterloo\\
    Waterloo ON L8S 3G1, Canada; glharris@astro.uwaterloo.ca}

\author{William~E.~Harris}
\affil{Department of Physics and Astronomy, McMaster University \\
      Hamilton ON L8S 4M1, Canada; harris@physics.mcmaster.ca}

\and

\author{Gregory~B.~Poole}
\affil{Department of Physics, University of Waterloo \\
     Waterloo, ON N2L 3G1, Canada; gbpoole@astro.uwaterloo.ca}

\begin{abstract}
We have used the {\it Hubble Space Telescope} to obtain 
WFPC2 $(V,I)$ photometry for stars in the halo of 
NGC 5128, the nearest giant elliptical galaxy.  The resulting
color-magnitude diagram (CMD) of this field, which
lies $\sim 21$ kpc from the center of the
galaxy, contains more than 10,000 stars and reaches almost 3 magnitudes 
down the red-giant branch (RGB).  From the sharply defined 
RGB tip at $I = 24.1 \pm 0.1$ and $M_I$(tip) $= -4.1$,
we obtain a distance to NGC 5128 of 3.9 Mpc.
Comparison with the fiducial RGBs of Milky Way
globular clusters and model isochrones demonstrates
that this outer-halo population of NGC 5128 is
completely dominated by old stars, with an extremely broad
metallicity range extending
from the most metal-poor Galactic globulars at [Fe/H] $\ltsim -2$ up
to above-solar abundance.  The relative contribution from
any younger, bright asymptotic-branch component is negligible.
The shape of the metallicity distribution function (MDF), derived from the CMD
by interpolation within the isochrones,  
can be remarkably well matched by a simple two-component
model of closed-box chemical enrichment, where the first component
starts with an initial abundance $Z_0 =0$ and the second component
with $Z_0 \simeq 0.18 Z_{\odot}$.  Two-thirds of the stars belong
to the metal-richer component and one-third to the metal-poorer one;
the mean metallicity of the
entire sample is $\langle$[Fe/H]$\rangle = -0.41$, consistent with the
colors of the integrated halo light.
The metal-rich component also coincides strikingly in mean
and dispersion with the metal-rich peak of the 
halo {\it globular clusters} in NGC 5128, suggesting that both of these
halo subsystems formed contemporaneously.  A discussion of various models of
E galaxy formation leads us to suggest that a basic {\it in situ}
formation picture with two distinct epochs of star formation
best fits the observations; other models involving major contributions from 
accretions or mergers are less satisfactory.  The timing of the events
we suggest is that the first, more metal-poor star-forming epoch took
place while the protogalaxy was still in a clumpy, fragmented state,
leaving most of the gas unused.  The second and larger star
formation epoch took place after the majority of the now pre-enriched gas
had recollected into the fully formed potential well of the
new giant elliptical.

\end{abstract}


\section{Introduction}

A color-magnitude diagram (CMD) for the stars in a galaxy is the definitive
tool to establish the nature of its stellar population, and even a CMD limited to
the brightest stars can provide considerable information on the
nature of its star formation history.  Such work is, however, difficult to
carry out for the halos of galaxies because the old-halo stars are fainter
than any Population I component, and can be directly resolved only for
nearby galaxies.  Thus for the most part, our
knowledge of galaxy halos has been based 
on surface photometry of their integrated
light.  Particularly important exceptions are, of course, the
Local Group members, notably M31 (e.g. \cite{mou86,pri88,chr91,dur94,rei98}) 
and the dwarf ellipticals (\cite{gri96,han97,mar98}).
Recently, direct CMDs have been obtained from HST imaging of the halos of
a small number of galaxies beyond the Local Group
(\cite{sor96,els97,cal98}).  

What is missing entirely from the Local Group is the
important giant elliptical class. 
Among all gE's, the closest one by far is
NGC 5128, the dominant galaxy in the Centaurus group at $d \simeq 3.9$ Mpc
(The next nearest gE target is NGC 3379 in the Leo group, more
than 2 magnitudes further away; while the rich collections of ellipticals
in Virgo and Fornax are more than 3 magnitudes more distant).
Although its central dust lane and gas, thought to be due
to the recent accretion of a smaller satellite, have traditionally 
marked NGC 5128 as `peculiar', its much larger main body and extensive outer halo
resemble those of more nearly normal gE's in structure, size, and color
(\cite{hui95,sor96,gra79,ebn83,vdb76,sto97}).  Since mergers and accretions
of the type now going on in NGC 5128 are increasingly recognized as rather
normal episodes in the histories of large galaxies, it is not unreasonable
to expect that we can learn a great deal about the stellar populations
in E galaxies by studying it intensively.

Though NGC 5128 is well beyond the Local Group, its halo red giant stars
are clearly within reach of the Hubble Space Telescope cameras.
To date, the only CMD study for the old-halo population of NGC 5128 
is the work of Soria et al.~(1996).  But their 
CMD for a halo field about 10 kpc
south of the center of NGC 5128 reaches only about 1.5 mag down the
red-giant branch (RGB), and (as will be seen below) gives an
incomplete picture of the full stellar distribution.

\section{Observations}

During HST Cycle 5 we obtained deep images of an outer-halo field
in NGC 5128, located at a projected distance $R_{gc} = 18\farcm 32$ 
south of the galaxy center (shown in Figure 1), equivalent to 
a linear distance of $20.8 (d/3.9)$ kpc.
For this program (GO-5905), ten orbits were
assigned from the Cycle 5 competition in mid-1994, and the observations 
were executed three years later, in two visits on 1997 August 16 and 24.  
We used the WFPC2 camera in the two filters F606W ($V$) and F814W ($I$).
Total exposures in each filter were 12800 sec 
($2 \times 1200$ sec $+ 8 \times 1300$
sec).  The PC1 chip was centered on the halo globular cluster 
C44 (\cite{hghh92,hes86}), as shown in Figure 2.

From the PC1 photometry we derived a CMD for the globular
cluster, the first one for a globular cluster in 
any galaxy outside the Local Group; these results are described in
Harris et al.~(1998a).  In the present paper, 
we discuss the main set of data from the three WF chips and the
CMD for the halo stars.

Starting from the preprocessed images supplied by StScI, we reregistered
all the individual exposures in each filter to a common coordinate system
and combined them with the normal {\it stsdas.crrej} package.
The exposures were sub-pixel shifted to facilitate the
removal of cosmic rays, bad pixels, and other artifacts on the frames.
Photometry on the summed $V$ and $I$ frames was then carried out with
the DAOPHOT II code (\cite{ste92}). We defined the point spread function
(PSF) using several moderately bright,
isolated stars in each WF field, and carried out two iterations of the 
normal sequence of object finding and PSF fitting through ALLSTAR.
Transformation of the instrumental magnitudes to $V,I$ followed the
prescriptions of Holtzman et al.~(1995).  No 
correction was made for the charge-transfer ramp effect across 
the frames (\cite{hol95,whi97}) because of the length of our exposures.  
The final CMD in $(I,V-I)$, for 10,352 
stars over the three WF fields combined, 
is shown in Figure 3a.  Objects with goodness of fit $\chi > 2$ are
excluded.

Artificial-star tests were then carried out to investigate the completeness
of detection and the internal measurement uncertainty as functions of
magnitude and color.
The ADDSTAR code was used to add $\sim 200$ identical stars at a time
to the WF2 frame for each color at 0.5-mag intervals across the CMD,
and over a 7-magnitude range in $I$.
The WF2 frame was chosen for the ADDSTAR tests because
it had the smallest number of very bright stars and background galaxies.
The ``error bars'' (internal measurement precisions $\sigma_I, \sigma_{V-I}$)
resulting from these tests are shown in Figure 3b, along with the 90\% and 50\%
detection completeness lines suitably transformed from the instrumental
$(v,i)$ magnitudes to the $(I,V-I)$ plane.
Above the f=0.9 line, the data have photometric uncertainties 
$\sigma_I\ltsim<$0.05 mag and $\sigma_{V-I}\ltsim$0.1 mag in $V-I$.  Between
the f=0.9 and 0.5 lines the data still have modest
uncertainties $\ltsim$0.2 mag in $I$ and $\ltsim$0.3 mag in $V-I$. 
The internal photometric errors increase very rapidly for completeness
levels less than $f=0.5$; in the following discussion, we will use 
none of the data fainter than this line, and almost none fainter than
the $f=0.9$ line.  

A complete electronic file of the final photometry may be obtained
on request from the first author.

% Authors may indicate to the editorial staff where they would like 
% figures and tables to be placed in the manuscript.  This is done with
% either the \placefigure{KEY} or \placetable{KEY} commands.  These
% commands require \label{KEY} commands to be placed appropriately with
% corresponding table and figure captions.  When the manuscript is
% printed a short note is printed on the page where the figure or table
% is to go.  These commands are ignored in the aaspp4 and aas2pp4 styles.

%\placetable{tbl-3}
%\placefigure{fig1}

\section{The Color-Magnitude Diagram}

The CMD in Figure 3a has several features that are immediately apparent.
(1) The number of stars rises abruptly at $I \simeq 24.1$, 
which (like \cite{sor96})
we interpret as the tip of the first-ascent red giant branch (TRGB).
The value of $I_{TRGB}$ has been shown to be a reliable distance
indicator for old metal-poor stars (cf.~\cite{lee93,mad95}), and will be
used in \S4 below for a new calibration of the distance to NGC 5128.
The following discussion will assume a  
distance of 3.9 Mpc, a foreground reddening of $E(V-I) = 0.14$, 
and thus an apparent distance modulus of $(m-M)_V = 28.2$.
(2) The fact that the CMD contains almost no objects
brighter than $I\sim 24$ indicates that the halo at this location
is dominated by an old stellar population, with little contamination
from crowded stars or highly evolved bright AGB stars (e.g., \cite{ren98}).
(3) Over the $\ltsim$3 magnitude range in $I$ of our reliable data, 
the range in $V-I$ color is extremely broad -- larger than has been seen 
in any other galaxy for which comparable data are available.  

A primary goal of our study is to obtain the star-by-star 
{\it metallicity distribution function} (MDF) from the CMD. 
Giant ellipticals are known to have much higher mean metallicities
than dwarfs or the halo of the Milky Way (e.g., \cite{bro91,kor89,tho89}), 
extending up
to solar [Fe/H] and perhaps even higher.  But
what is the actual shape of the MDF which produces this high mean?
As a first indication of the range of metallicities across the RGB,
in Figure 4 we have superimposed the mean lines for standard
Milky Way globular clusters of different metallicities
(\cite{dac90} and \cite{gua98}), shifted by our adopted reddening
and distance modulus given above.  
These lines are for M15 ([Fe/H]$=-2.2$), NGC 1851 {[Fe/H]$=-1.3$), 47 Tuc 
([Fe/H]$=-0.7$), and NGC 6553 ([Fe/H]$= -0.25$). 
Although the photometry for NGC 6553 (\cite{gua98}) is less
precise and the locus less well defined than for the other clusters
because of its crowding and high
reddening ($E(B-V)= 0.7$), it is the most metal-rich globular
cluster which we can call on with some level of confidence.  

From the comparison in Figure 4, it is clear that the RGB stars extend
fully from [Fe/H]$\sim -2.0$ to at least solar abundance and perhaps higher.
A similar but not so obvious pattern appears in the CMD
of Soria et al (1996; see their Figure 3).  Blueward of the NGC 6397
([Fe/H]$=-2.0$) locus, the number of bright giants diminishes markedly 
in their CMD but does not drop to zero.
In our CMD, many stars fainter than $I \sim 26$ fall to the blue side of the
M15 line; we interpret these as due to a combination of photometric scatter
(Figure 3b) and the presence of AGB stars, whose tracks lie $\sim 0.1$ mag
blueward of the RGB in this magnitude range.

We also compare our data with stellar models in Figure 5, using 
isochrones from Bertelli et al.~(1994).  More or less
arbitrarily, we adopt their models for an age
of 12.5 Gyr, and show the lines for
their full range of metallicities, [Fe/H]$= -1.6$ to
+0.47.  Comparing these isochrones with the fiducial 
globular cluster lines in Figure 4,
we find that the isochrones at [Fe/H] $\simeq -1.2, -0.6,$ and $-0.3$ 
match remarkably well with the fiducial cluster lines of similar metallicity.
The [Fe/H]$=-1.6$ isochrone, in turn, matches well with M15 
even though it is nominally 0.6
dex more metal-rich; we will not speculate here on the reasons for this.
For the two most metal-rich isochrones at [Fe/H]=+0.07 and +0.47 we 
unfortunately have no direct observational comparison with real globular
clusters.  Observational scatter aside, the main conclusion to be
drawn from Figures 4 and 5 is that this
NGC 5128 halo field is dominated by an old 
stellar population with an extremely broad range in heavy
element abundance, possibly exceeding 2.0 dex.

It is also possible that star-to-star age differences
may exist, and be responsible for some of the color spread.  
To explore the effect of age, in Figure 6 we compare the CMD with
isochrones of various ages for a single metallicity, [Fe/H]$= -0.32$ 
(similar to the mean metallicity for the entire sample; see \S6 below).
This comparison demonstrates the conventional result
that the color of the old RGB is affected much more strongly by 
metallicity than age; the observed color range in our data is far too great
to be explained by a large range in age combined with
a small or even moderate range in [Fe/H].  We conclude that the broad color
range we observe must be due primarily to its range in chemical
composition.  This will be exploited further in \S6 to derive the MDF.

\section{The Brightest Stars and the Distance Calibration}

The absolute magnitude of the RGB tip is well established as an
excellent standard candle for stars with [Fe/H]$\ltsim -0.7$. 
Calibrations of the absolute $I$ magnitude of
the TRGB adjusted for recent Hipparcos
parallax data give $M_I$(tip) $= -4.1 \pm 0.1$
(\cite{mou86,lee93,mad95,sak96,gra97,hpds98}), which we adopt here.
In Figure 7, we show the luminosity function of our entire dataset,
as well as the LF for only those stars bluer than the [Fe/H] $= -0.6$ 
fiducial isochrone line in Figure~5.
A rapid rise at $I_{TRGB} = 24.1 \pm 0.1$ is evident in both
histograms. Although the calibration $M_I$(tip) $=-4.1$ applies strictly
to the more metal-poor stars, it can be used 
for the whole LF over all metallicities
just as well (knowing {\it a priori}, of course, that a 
significant population of low-metallicity stars is present).  
With $I_{TRGB} = 24.1 \pm 0.1$, $M_{I} = -4.1 \pm 0.1$, 
$E(B-V) =0.11$ (\cite{bur84}) and thus 
$E(V-I) = 0.14$, $A_I = 0.22$, we therefore calculate a distance modulus of
$(m-M)_0 = 27.98 \pm 0.15$, or $d = (3.9 \pm 0.3)$ Mpc.  
This estimate agrees well with the planetary nebula luminosity 
function value of $(m-M)_0 = 27.97 \pm 0.14$ (\cite{hui93}); we have 
adjusted their
published value upward by 0.2 mag to correct it to the contemporary Local
Group distance scale, $(m-M)_0$(M31) = 24.5 (\cite{vdb95,fer98,saas98}).
The surface brightness fluctuation method (\cite{ton90,hui93})
gives $(m-M)_0 = 27.78 \pm 0.10$, adjusted to the same M31 
distance as above.

The number of stars {\it brighter} in $I$ than the TRGB is very small:
there are $\simeq 50$ objects in the CMD with $22.0 < I < 24.0$, i.e.~about
0.5\% of the total population.  There will be four kinds of objects
contributing to the numbers above the tip:  

\noindent (a) Accidental blends of bright RGB stars (\cite{ren98}): 
using Renzini's prescriptions, we can quickly show that the
expected number of these blends is entirely 
negligible (there are $\sim 1000$ stars within about 1 mag of the
RGB tip, spread over $1.9 \times 10^6$ pixels, giving $N_{blend} < 1$).  

\noindent (b) Giants or AGB stars in highly
evolved states including the long-period-variable (LPV) or Mira states:
the expected number of these is proportional to the total luminosity
of the sampled stellar population.  The integrated magnitude of
all the stars in our sample brighter than $V = 27.5$ (roughly, the 
average completeness limit of the $V$ photometry) is $V_{int} \simeq 17.72$.
Using a normal Population II (globular cluster) luminosity function
to extrapolate the luminosity function
to fainter magnitude, we estimate very roughly that 
the total luminosity over all magnitudes is then 
$V_{int} \simeq 16.1$, equivalent to $L_{int} \simeq 5.9 
\times 10^6 L_{\odot}$.  Again using Renzini's (1998) scaling relations,
we would then expect to find $\sim 10$ LPV-type stars generated from the
old-halo population within our field.

\noindent (c) Foreground field stars, and any ``starlike''
background objects such as QSOs: over our small sampled area of
5.33 arcmin$^2$, the number of objects in the latter category
will be negligible.  However, the predicted number of
foreground stars in the range $I = 22 - 24$ (e.g., \cite{bah81})
is $\simeq 5$ per arcmin$^2$, which yields about 25 expected
bright foreground stars across our field.  This makes up about
half the number of bright stars we see.

\noindent (d) Any genuinely younger population of AGB stars:
by process of elimination, we need to ascribe at most about 15 objects
to this category; 
the other $\sim 35$ stars brighter than the RGB tip are accounted for
as either LPVs or foreground stars.

We conclude that this part of the NGC 5128 halo is
almost purely an ``old'' sample
of stars; whatever recent star formation has affected the 
inner few kpc of the galaxy has not penetrated outward to contaminate 
this part of the halo.

\section{Comparison With Other Galaxies}

Most of the direct color-magnitude data currently available 
on halo populations in other
galaxies is for Local Group systems such as M31
(\cite{mou86,pri88,dav93,arm93,dur94,rei98}), M33
(\cite{mig95,mou86}), and the dwarf ellipticals including
NGC 147 (\cite{han97}), M32 (\cite{gri96}), and NGC 185 (\cite{mar98}).
With the HST cameras, such efforts have extended beyond the Local
Group to include the previously discussed short-exposure study of 
NGC 5128 (\cite{sor96}), two dwarf spheroidal 
companions of M81 (\cite{cal98}), and the S0 galaxy NGC 3115
(\cite{els97}).  But, with the possible exception of NGC 3115, no other
galaxy halo shows the same range of heavy element enrichment that
we see in NGC 5128 amongst its old stars.

In Figure 8 we have plotted $I,(V-I)$ CMDs for our data along with
three of the small ellipticals, NGC 147 and the two M81 dwarfs.  Marked
on each graph are the number of objects plotted and the fiducial lines for
the most metal-poor and metal-rich globular clusters (M15 and NGC 6553).
We note first, of course, that the NGC 5128 data exhibit by far the
greatest color span of any of the four.  Second, in relative
numbers, NGC 5128 has the fewest stars {\it above} the TRGB.
NGC 147 and M81-BK5N are quite similar to NGC 5128 in the location of the blue
edge of the CMD, suggesting their original halo gas started with nearly
primordial compositions; NGC 5128, however, has gone very much farther up
the enrichment ladder.  The CMD for the halo of NGC 5128 {\it is the first
clear demonstration that giant E galaxies can possess extremely low-metallicity
stars} as well as high-metallicity ones.  

The only other galaxy mentioned above which may show
a comparably broad abundance distribution  
is the S0 galaxy, NGC 3115 (\cite{els97}).  But NGC 3115 is more than
2 magnitudes more distant, with $I_{TRGB}\sim 26.0$.  
Consequently, the CMD in Elson's study covers only the top part of the RGB,
where the evolutionary tracks spread out rapidly with metallicity and
where it is hardest to deduce the actual metallicity range and MDF.
An additional factor hampering the interpretation of the NGC 3115 data
is that its exposure time in $V$ was only $\sim$2/3 of that in $I$
and, consequently, the very reddest stars ($V-I > 3$) have much more
uncertain colors or are cut off entirely by the 
photometric incompleteness limit (see Figure~3 once again).
The same was true to a somewhat lesser degree for the Soria et al.~
study of an inner-halo field in NGC 5128.  Since their results cover only
the top $\sim 1.5$ mag of the giant branch, their CMD clearly shows
the metal-poor population, but (as they discuss), 
they can only place lower limits on the numbers of reddest stars.  
Even in our data, which cover $\sim 3$ magnitudes of the RGB, the very
reddest stars fall below the $f=0.9$ completeness line (Figure 3).
These extremely red stars can be fully surveyed 
only by increased exposure time in the bluer filter.

It is intriguing to consider the
possibility that an S0 galaxy (NGC 3115) might have experienced a heavy
element enrichment and star formation history similar to 
that of a gE such as NGC 5128.
Unfortunately, given the distance of NGC 3115 and current observational
limitations, it is unlikely that we will be able to obtain data of
the necessary quality for a rigorous comparison -- 
or the giant ellipticals in Leo, Virgo, and Fornax -- in the near future.
 
\section{The Metallicity Distribution Function}

A closer look at the color distribution is shown in Figure 9, where
we have plotted histograms in $V-I$ for five magnitude intervals
from $I=24.0$ to 26.5.  The distribution is broadest at 
the top end where the evolutionary tracks are most separated,
but it narrows in the lower bins to the point
where the two faintest samples are strongly peaked.  
For NGC 3115, Elson (1997) suggests that the color distribution
of the brightest halo stars is bimodal.  In NGC 5128, we see no
clear evidence for bi- or multi-modality in any of the color
distributions; this does not, however, mean that the {\it metallicity}
distributions are necessarily unimodal.

To derive the full MDF, we need to convert $(V-I)$ to [Fe/H] on a
star-by-star basis.  The shape of the color histograms (Figure 9) does
not give a correct indication of the MDF, since the conversion from
color to metallicity is highly nonlinear, as is obvious in Figure 4.
Ideally, to estimate metallicities 
we would prefer to use the fiducial tracks for the Milky Way
globular clusters, which are on a well established
and observationally based metallicity scale.  
However, most of the stars in NGC 5128
are clearly more metal-rich than our calibrators (47 Tuc, and even NGC 6553).
Similarly, the observationally based conversion relations of 
Da Costa \& Armandroff (1990) are valid only for [Fe/H]$\ltsim -0.7$ and thus
irrelevant for NGC 5128.  

To cover the full color range necessary, we have used a combination
of the real cluster fiducial lines (all the ones shown in Figure 4)
and the two most metal-rich isochrone lines shown in Figure 5.
Although we regard this only as a preliminary procedure,
the fact that the lower-metallicity isochrones agree closely in color
with the globular cluster lines (noted above) gives some 
confidence that systematic errors in [Fe/H] will not be serious.

In the brightest $\sim 1$ mag of the RGB, the tracks have 
strong curvature and interpolation between them is extremely sensitive
to small errors in color.  We have therefore chosen
simply to avoid this region of the CMD, which in any event contains only
a minority of the stars.  Instead, we use only the data points further down
the giant branch in the range $25.0 \ltsim I \ltsim 26.5$ where
the tracks are more nearly vertical and interpolation between them
is more reliable.  For each star in this range, we then
(a) determined the $(V-I)$ color of each fiducial line
at the $I$ magnitude of the star, and (b) interpolated in the
($(V-I)$, [Fe/H]) relation at that magnitude to predict [Fe/H] for
the given star.\footnote{Over the full range of metallicities shown here,
the conversion is highly nonlinear. Simple linear interpolation 
between $(V-I)$ and [Fe/H] within any pair of the fiducial lines produces
systematic errors which show up as artificial ``blocks'' of stars in
the [Fe/H] histogram.  To circumvent this,
for each star we first calculated a
hyperbola-like function of color, $C = ((V-I) - (V-I)_{min} + 1)^{-2.6}$
where $(V-I)_{min}$ is the color of the bluest fiducial line at the given $I$
magnitude.  Numerical trials showed that this $C$ index was nearly a linear
function of [Fe/H] over our magnitude range of interest, allowing interpolation
which was much more accurate and which did not generate artifacts in
the MDF.  This index was used
purely for numerical convenience and is not suggested to have any physical
significance.}  A very similar procedure was used by Grillmair et
al.~(1996) in their study
of M32 to develop a rough histogram of metallicity from the CMD.

The resulting metallicity distribution is displayed in Figure 10,
for three 0.5-magnitude ranges along the RGB.
All three plots show a broad, shallowly sloped distribution at low
metallicity, then a sharp rise near [Fe/H]$= -0.6$ to a peak near 
$\simeq -0.3$. Very few stars lie above solar metallicity.
The overall shape of this distribution cannot be easily described as
``bimodal'' in the conventional sense, though it clearly has at least
two major components.  
The approximate internal uncertainties in [Fe/H] (shown in the 
top panel, for $I \simeq 25.5$)
are, as expected, greatest at low metallicity where the isochrones 
are closest in color for a given $\Delta$[Fe/H].  Fortunately,
this low-metallicity region of the MDF is least affected by the random
photometric scatter because of its shallow slope; and 
in the more rapidly changing high-metallicity component, 
the photometric scatter has an intrinsically smaller effect.  
Before we continue, we need to consider
two potential sources of systematic error in the 
overall shape of the distribution:

\noindent (a) Asymptotic-branch (AGB) stars must be present along with
the first-ascent red giants that we are interested in.  Since they are
always bluer at a given metallicity, they will cause us to overestimate
the relative numbers at the low-metallicity end.  At the luminosity levels 
represented here (1 to 2 mag below the TRGB), the AGB tracks are 
$\simeq 0.05 - 0.10$ bluer in $(V-I)$ than the RGB (\cite{ber94}).
Inspection of the lifetimes along the evolutionary tracks shows that,
over the range $25 < I < 26$, about $(20 \pm 3)$\% of our total 
sample will be AGBs stars, nearly independent of metallicity.
Their presence will therefore not dramatically skew the MDF (see below).

\noindent (b) As was noted earlier, a bias against the relative numbers of stars
at the extreme red end is introduced by the photometric completeness line, which
slants upward through the red tip of the RGB (Figure 3).  We have already 
argued that this bias is considerably smaller in our data than in 
Soria et al.~(1996) or in Elson (1997) because we reach much further down the
giant branch.  In addition, the fact that
there is a clear gap in the CMD between the red edge of the RGB population
and (e.g.) the 50\% completeness limit in Figure 3 suggests that there are
genuinely few stars more metal-rich than approximately solar abundance.
However, close comparison with the isochrone lines suggests that we will be slightly
underestimating the {\it relative} numbers of stars more metal-rich than
[Fe/H] $\simeq -0.2$.  Correcting for this would extend the upper envelope 
of the MDF in Figure 10 but would not change the peak location.

For the present, we have not attempted to correct for either of these
minor effects quantitatively, and we repeat that we consider
the MDF derived here to be only a preliminary one.
Ultimately, a better procedure for extracting the
[Fe/H] distribution would be to use a finer grid of fiducial RGB and AGB
lines and an assumed input MDF to synthesize the observed color-magnitude
array, appropriately convolved with the photometric error and completeness 
functions (cf. \cite{mar98} for an example of this technique).
We are currently exploring this approach. 

The general shape of the MDF in Figure 10 is quite similar in all three
magnitude bins, indicating that the
$(V-I)$ to [Fe/H] conversion was internally consistent.
The faintest of the three magnitude bins ($I=26.0-26.5$)
is, however, clearly broader than the two brighter ones, a direct result of
the increased random photometric errors there.  For the purposes of the
present discussion, we exclude this faintest
bin, combining only the two brighter ones in which the internal
photometric scatter does not exert an important effect on the
internal spread of the MDF.  The resulting composite MDF
of 4514 stars is shown in Figure 11 (upper panel).
The ``metal-rich'' component ([Fe/H] $\gtsim -0.7$) contains
$\simeq 64\%$ of the stars in the distribution.  This dominant
section of the MDF is reasonably matched by a simple Gaussian curve
(shown in Figure 11) with a narrow dispersion $\sigma$[Fe/H] $\simeq 0.22$
and a maximum at [Fe/H]$_{peak} \simeq -0.32$.

At lower metallicities, we see the very extended and less populous
component (36\% of the total) sloping away to [Fe/H] $\sim -2$ and below.
The upper part of this section (from [Fe/H] $\sim -1.1$ to $-0.7$)
will be slightly contaminated by AGB stars from the metal-rich component
itself; taking into account the behavior of the
translation from $(V-I)$ to [Fe/H],
we estimate very roughly that 50 to 100 of the $\sim 600$
stars in this intermediate [Fe/H] range will be AGB stars that
have bin-jumped from the higher-metallicity component.  Thus 
in relative terms, 
the histogram near [Fe/H] $\sim -1$ is about 10\% higher
than it should be.  This offset will not affect any of the following
discussion critically.

The mean abundance of the entire sample of stars, combining
both components, is $\langle Z \rangle = (0.39 \pm 0.02) Z_{\odot}$
or $\langle$[Fe/H]$\rangle = -0.41 \pm 0.03$. 
Adjustment for the second-order
bias corrections mentioned above would raise this mean even higher,
perhaps by $\sim 0.1$ dex.   This metallicity level
is high for such a remote halo location, but is not inconsistent
with the integrated colors and 
spectral parameters for E galaxies in general, which usually
yield rather modest radial metallicity gradients and mean values in
the range of [Fe/H] $\sim -0.4$ to above-solar (e.g. \cite{tho89,kor89}).  
Thus, our data strongly support the view that
the problem of producing high-metallicity stars at extreme
outer-halo distances is a generic one for ellipticals.

The only surface photometry for NGC 5128 extending to large radii
($r \simeq 10'$) is from van den Bergh (1976), based on $UBV$ photoelectric 
photometry.  Although his outermost points are at only half the 
projected radius of our field, they provide a useful comparison.  
Using the conversion relation of \cite{cou90} to translate integrated
color roughly to metallicity,
$(B-V)_0 = 0.200$[Fe/H] + 0.971, we find [Fe/H] $\sim 0.0 \pm 0.5$
for fields $8'-10'$ from the center, in satisfactory agreement 
with the much more accurate mean metallicity we have
calculated directly from the MDF.  

Lastly, in Figure 12, we have replotted the MDF 
on a linear scale (number per unit abundance $Z/Z_{\odot}$,
where we adopt $Z_{\odot} = 0.017$ as the solar abundance).
In this graph, the dominance of the metal-richer population,
and the small true range in abundance of the metal-poor component,
are more clearly visible.

\section{Discussion:  Implications for Galaxy Evolution}

The full MDF undoubtedly conceals many details of the enrichment 
history for NGC 5128 and, at this stage, we can provide 
only a rough interpretation.  However, the overall shape of the MDF 
already indicates that a monolithic, single-burst
formation event is unlikely to be an appropriate scenario for this galaxy.

\subsection{Self-Contained Formation Scenarios}

We first attempt to interpret the MDF in the context
of an {\it in situ} formation picture:  that is, we suppose that 
the galaxy was built primarily out of gas within its original 
protogalactic potential well, without important later additions 
from outside.  Of course, this assumption cannot be strictly correct, since
an accretion event with infalling gas and recent star formation is
going on now, and other such events may have happened previously.
Nevertheless, the general shape of the MDF can be matched surprisingly
well by a combination of just two major star-forming phases:
an initial one producing about one-third of the
stars, starting from primordial (essentially unenriched) gas;
and a later, more prominent phase producing two-thirds of
the halo and with its maximum contribution at $Z \sim 0.5 Z_{\odot}$.

A quantitative match to the MDF is shown in Figure 13, 
based on the classic picture of 
closed-box, one-zone chemical enrichment
(e.g., \cite{pag75}).  In these models the cumulative distribution 
of stellar abundance $Z$ is given by
$$ N(Z) \sim {\rm const} \, (1 - e^{-(Z-Z_0)/y}) $$
where $N$ is the number of stars with heavy-element abundance less than $Z$;
the initial (pre-enriched) abundance of the gas is $Z_0$; and the yield rate
is $y$.  In Figure 13, the model curve for the
lower-metallicity component has $(Z_0, y) = (0, 0.002)$,
while the same quantities for the higher-metallicity component are
$(0.003, 0.005)$.  Although these particular parameters are only
illustrative -- other combinations with roughly similar values 
can be found which give similarly accurate fits -- 
the overall quality of the match to the data is remarkable
even with these very basic assumptions.

Within the context of this model, we would conclude
that the first round of star formation proceeded at rather low global
efficiency and low yield, leaving most of the gas unused till later.
Exactly when and how these two epochs took place is still a matter of
speculation (the MDF itself gives little direct information 
on timescales), but contemporary galaxy formation modelling 
may provide some guidance.  In standard cold dark matter cosmologies,
large galaxies are generally thought to have formed 
from heirarchical merging of dwarf-sized clumps which 
appear naturally within N-body simulations of galaxy formation
(e.g., \cite{fre88,nav91,lac94,qui96,nav97,wei98}, among
many others) and which are also postulated on 
observational grounds to be the sites of early 
globular cluster formation (\cite{sz78,hp94}). 
We can then imagine that the first burst of 
star formation took place within these clumps, 
starting at very low metallicity and enriching
the residual gas in the still-lumpy potential well 
of the proto-elliptical up to an intermediate
level near [Fe/H] $\sim -0.7$.  The first wave of 
supernovae and massive stars would have ejected much of the
gas from the small local potential wells of
these dwarf-sized fragments before it had a chance to
form many stars.  Later, as the partially enriched gas
recollected in the now-complete galactic potential well,
a second and more major star-forming epoch
took place which generated the main stellar bulk of 
NGC 5128 and enriched it to near-solar abundance.  

These two early phases laid down the dominant
features of the halo that we now see.
Later on, the galaxy may have built further by accretion (with one such
event going on now), but these more recent events have plainly been
most important in the inner few kpc, with the outer halo 
largely unaffected by them.

Interestingly, Grillmair et al.~(1996) 
have found roughly similar features in the
halo population of the small Local Group elliptical M32.  Its MDF shows
a metal-rich peak at [Fe/H] $\simeq -0.2$ very much like NGC 5128,
but with an overall distribution that is somewhat narrower.  
The ``tail'' extending
to lower [Fe/H] is present in M32, but is less populous
and does not seem to extend below about
[Fe/H] $\sim -1.5$.  Just as we find here, 
Grillmair \etal\ note that a single one-zone model
does not fit the total MDF; a combination of at least
two separate enrichment events is necessary, or some more complex
history.  Although NGC 5128 and M32 by no means represent the full range of
ellipticals, it may not be premature to treat them as
the first examples of a general pattern for such galaxies.

\subsection{Connections with the Globular Cluster System}

An additional important piece of evidence is shown in 
Figure 11b.  Here, we show the MDF for the {\it globular clusters} in the
halo of NGC 5128 (\cite{hghh92}; the cluster metallicity measurements are
based on the sensitive Washington indices, principally $(C-T_1)$).
This sample comprises 47 clusters with projected distances 
$4' < r < 22'$ from 
the center of NGC 5128; although they are located all over the halo,
the globular cluster system has no significant metallicity gradient outside
the central region (\cite{hghh92}), so the MDF shown here is quite
representative of the clusters anywhere in the halo.
Figure 11b displays the well known bimodal shape of the MDF seen in the
globular clusters of several
giant ellipticals as well as large disk galaxies.  For our purposes,
a remarkable feature of this plot is that the MDF of
the {\it metal-rich} clusters {\it exactly coincides in peak location and
dispersion} with the metal-rich stars, to well within the observational
uncertainties.  We suggest from this evidence that the metal-richer
stars and globular clusters formed together, and that both represent the same
generic population that makes up the main bulk of the galaxy.

Still another remarkable point of
comparison is that for the clusters, the metal-richer 
component makes up only $\simeq 40\%$ of the total cluster
population, while the metal-poor component is 60\% of
the total.  These proportions are
almost the reverse of the numbers for the halo {\it stars}.
{\it If} we can safely associate the metal-poor clusters with 
the metal-poor stellar
component (even though their MDFs have rather different shapes), 
these ratios imply that the {\it globular
cluster specific frequency} $S_N$ (the number of globular clusters per
unit galaxy luminosity; \cite{wha91}) is {\it 
$\simeq 2.8$ times higher in the metal-poor component than in the metal-rich
component}.  Using the fact that the global specific frequency for the
galaxy is $S_N(tot) = 2.6 \pm 0.6$ (\cite{har98}), we can then derive
specific frequencies separately for the metal-poor and 
metal-rich populations.  These are
$S_N(MPP) \simeq 4.3$ and $S_N(MRP) \simeq 1.5$.  
The former value is similar
to the more cluster-rich ellipticals in Virgo and Fornax, while the latter
is closer to the values for disk galaxies (\cite{wha91}).

A similar calculation has been done for
the Virgo gE NGC 4472, where $S_N$ for the metal-poor
population is estimated to be more than an order of magnitude higher than for the
metal-rich population (\cite{gei98,for96,saas98}).  Within the context of
{\it in situ} formation, we can speculate that the clusters 
formed very early on in the first round of star formation,
but that most of the gas which could have formed stars throughout the halo
remained unused till the second round, leaving a low yield and high $S_N$
as its distinctive traces.  
Harris \etal\ (1998c) have suggested that
one mechanism which may have prevented the 
low-metallicity round of star formation
from running to completion was the first burst of supernovae, 
igniting early enough that the proto-halo still had low overall density.  
We might also speculate that a high fraction of the gas had not yet
collected into the dwarf-sized fragments within which star and cluster
formation is believed to take place most efficiently.

We note that the formation picture we have
just described -- a two-phase {\it in situ} model with later minor 
effects from accretion and merger -- is quite similar to the one
developed by Forbes et al.~(1997) from
a comparison of the globular cluster systems in E galaxies.
The MDF for the halo {\it stars} in this particular E galaxy provides
a powerful new consistency test for this scenario.

\subsection{Constraints on Accretions and Mergers}

Can these new observations for NGC 5128 fit equally well into
other formation scenarios?
An interesting alternate picture for E galaxies is that of
\cite{cot98}, who suggest that the high-metallicity component 
was the main part of the original galaxy,
while the low-metallicity component was
acquired later on by a series of accretions of smaller satellites.
Since smaller galaxies are more metal-poor, the relative number
of low-metallicity stars would therefore increase with time,
as long as the main process of accretion is ``passive'' (i.e., with
low gas content and no new formation of stars).
For NGC 5128, the accreted material would
need to make up as much as one-third of the total galaxy, 
equivalent to about 200 dE galaxies 
averaging $M_V \simeq -15$ each.  NGC 5128 is by
far the largest galaxy in the sparse Centaurus group, so these numbers
would require that it has already swallowed the great majority
of the dwarf population that originally existed in the group.
Currently the Centaurus group has about 20 dwarfs, mostly
irregulars (\cite{cot97}).
Although accretion is obviously playing an ongoing role in the history
of NGC 5128 at some level, these total numbers seem uncomfortably large
if the entire metal-poor component of the
MDF is to be explained this way.  In addition, the specific
frequency for the metal-poor component is like that of normal cluster-rich
dwarf ellipticals (\cite{dur96}) rather than cluster-poor dwarf
irregulars, so we would have to postulate that the metal-poor component
of NGC 5128 was built up mostly by accreting dE's.  
But dwarf irregulars vastly outnumber dE's in most small
galaxy groups like Centaurus. Thus {\it a priori} it is much more probable 
that NGC 5128 would have built up a metal-poor 
halo with {\it low} specific frequency, contrary to what we see.

Finally, it is not clear that a combination of many dwarfs of 
various sizes and individual enrichment histories would produce an MDF with
the simple shape that we see (basically well matched by a single 
enrichment event starting from primordial composition).
It is perhaps noteworthy that the MDF for the
Milky Way halo has a roughly symmetric shape centered on [Fe/H] $\sim
-1.7$ (e.g., \cite{lai88}), rather unlike the metal-poor component in
NGC 5128.

A possible counterargument to the rather simple calculation
just outlined above is that it may seriously
overestimate the fraction of the {\it whole} galaxy made up
by the metal-poor component.  Our one target field is located well
out in the halo, where presumably the relative numbers of metal-poor
stars are largest.  If we were able to sample the MDF over all
radii, the ratio of metal-rich to metal-poor components
$N(MRC)/N(MPC)$ might reasonably be expected to
increase inward; or, at smaller radii ($r \ltsim 5$ kpc, the region
most affected by later accretions of gas-rich satellites), we might
expect to see the emergence of a third and still more metal-rich component
that was built up through later star formation in the central
bulge.  However, in this case the specific frequency problem with
the globular cluster system would become even worse.
If, for example, the global value of $N(MRC)/N(MPC)$ is $\sim 0.1$
instead of $\sim 0.3$, then we would have $S_N(MPC) \simeq 12$,
bringing it up into the realm of the high$-S_N$ galaxies like
M87 and the central cD's in rich clusters (\cite{har98}).  
Passive accretion of dwarf galaxies, each of which has $S_N \sim
1-6$, seems highly unlikely to yield such a result.

More or less the opposite approach is taken in the merger model of
\cite{ash92}.  Here, the metal-rich component is assumed to form during
the merger of very gas-rich progenitor galaxies.  The merger-driven
star and cluster formation is postulated to occur predominantly in
the central regions where the gas collects most strongly, 
while the metal-poor
component represents the combined halos of the original pre-merger objects.
However, part of the \cite{ash92} picture is that the metal-rich 
component should have {\it high} specific frequency (i.e. that globular
clusters are postulated to form more efficiently in the colliding
gas clouds).  This is, unfortunately, exactly the opposite of what
we find in NGC 5128, where $S_N(MRC)$ is several times smaller than
in a typical gE.  In addition, we clearly find that the metal-rich component
represents {\it most} of the galaxy even far out in the halo, and not
just a minor addition.  In this view, we would have to assume that
the pre-merger ``galaxies'' were themselves almost entirely gaseous,
which in our view considerably blurs the distinction 
between a merger and {\it in situ} formation picture to begin with.

No single model is fully satisfactory or fully quantitative at present. 
Nevertheless, we believe that a plausible explanation of the data for
NGC 5128 is that the bulk of the
galaxy was built during two rather distinct star-forming epochs, from
an original supply of gas within its global potential well.  
Later accretions of satellites have probably made minor additions 
to the total mass, particularly in the central regions.
We also suggest that a better assessment of the various models will
be possible once we can sample the stellar MDF at several different places
in the halo and measure its radial gradient.

\section{Summary}

We have used WFPC2 photometry to derive a color-magnitude diagram for
the halo stars in NGC 5128, reaching almost 3 magnitudes down the
giant branch.  We find that the stellar population is completely
dominated by a classic old RGB population, with negligible contribution
from any younger bright AGB component.  The metallicity range of the
stars is extremely broad, extending from [Fe/H] $\ltsim -2$ to 
above-solar abundance.  The mean metallicity of the entire sample
is $\langle$[Fe/H]$\rangle = -0.41$.

From the CMD and the fiducial isochrone lines, we have generated
a first-order measurement of the 
metallicity distribution function for the RGB stars, the first
such data for any giant elliptical galaxy.
The MDF is well matched by a remarkably simple model
with two distinct phases of chemical enrichment.  In addition, we find
that the metal-richer component (which makes up two-thirds of all
the stars) has the same peak location and metallicity dispersion 
as the metal-richer {\it globular clusters} in the halo of NGC 5128,
suggesting strongly that both types of objects formed together.
The match between the more metal-poor stars and metal-poor globular
clusters is less exact.  However, the {\it specific frequencies}
of the two components are strikingly different:
the metal-poor component produced almost three times more globular 
clusters per unit halo light than the metal-rich component.

Comparing different formation scenarios for E galaxies ({\it in situ},
accretion, mergers) with the combined observations from the
halo stars and the globular clusters, we suggest that the most
satisfactory description is a two-phase {\it in situ} model, in which
an early, widely dispersed burst of star formation starting from 
primordial material left behind the metal-poor stellar component, 
but also left most of the original gas supply unused.  Then, a second 
and more major burst took place, giving rise to the metal-rich component
which forms the majority of the stars.  Models in which (for example)
the metal-poor component was accreted later, or in which the
metal-rich component was formed in mergers, are clearly less
satisfactory.

NGC 5128 is becoming a rich mine for stellar population
studies that will help us reconstruct its formation history.  
Further observational steps can easily be made which will add new
dimensions to its study:  for example,  additional 
HST photometry to the same depth that we have used here,
at different radial locations in the halo, would allow us to
measure the radial metallicity gradient and the progressive change
in the entire MDF.  Eventually, with more advanced imaging tools,
we can look forward to obtaining far larger samples of stars, 
penetrating further down the color-magnitude diagram to the 
horizontal branch and below, and finding fine structure in the
metallicity distribution function which will lead to more advanced
modelling.

\acknowledgments
This work was supported financially through the Natural Sciences
and Engineering Research Council of Canada, through research grants
to G.L.H.H.~and W.E.H.

\clearpage

% Option 3. Figures and figure captions are included within figure 
% environments within the body of the manuscript.  In our examples the 
% \plotone command is placed in the figure environment along with the
% figure caption.  The \caption command can also include a \label command.
% Each figure and its caption are printed on the same page.
%
% The \caption command in the figure environment works like the one in the
% table environment (it's the same one, actually), except that this one
% produces identification text that reads "Figure N."
%
% If you wish to see this option then you must comment out all of the 
% \figcaption, \plotone, and \end{document} commands above.

\clearpage

\begin{figure}
\caption{The south halo region of NGC 5128, 
showing the location of our WFPC2 field (labelled HHP) and the
field studied by Soria et al.\ (1996, labelled S96).
North is at top and East at left.
Our field is located at a projected distance $R_{gc} = 18\farcm32$ from
the center of NGC 5128, equivalent to 20.8 kpc for a galaxy
distance $d = 3.9$ Mpc. \label{fig1}}
\end{figure}

\begin{figure}
\caption{A mosaic of our HST WFPC2 field. Note the globular
cluster C44 (\protect\cite{gha98}) at the center of the PC1 chip.  
The dark swaths across the lower right of PC1 are scattered light from
a bright star off the edge of the field.  This image is the combined
sum of all 10 long exposures in $V$, totalling 12800 sec.
The data presented here are the results from the WF2,3,4 fields.
The directions (N,E) are only approximate; see the previous Figure.
\label{fig2}}
\end{figure}

\begin{figure}
\caption{{\it Upper panel:} The $I$ vs. $V-I$ color-magnitude diagram for 10352
measured stars with good PSF fits ($\chi < 2$).
The superimposed lines represent
$90 \%$ (dashed) and $ 50 \%$ (solid) photometric completeness levels,
as determined from artificial-star tests (see text).
{\it Lower panel:} Error bars showing the photometric measurement uncertainty as a 
function of magnitude and color.  
\label{fig3}}
\end{figure}

\begin{figure}
\caption{The color-magnitude diagram from Figure 3, with fiducial lines
of four Galactic globular clusters superimposed.  These are
M15 ([Fe/H]= $-2.2$), NGC 1851 ([Fe/H]=$-1.3$), 47 Tuc ([Fe/H] =$-0.7$) from
Da Costa \& Armandroff (1990), and NGC 6553 
([Fe/H]=$-0.25$) from Guarnieri et al.~(1998). The NGC 5128 halo stars cover
an extremely wide metallicity range.  \label{fig4}}
\end{figure}

\begin{figure}
\caption{The color-magnitude diagram from Figure 3, with theoretical
isochrone lines from \protect\cite{ber94} superimposed.  The isochrones have
an age of 12.5 Gyr and metallicities as labelled from $-1.6$ to +0.5. \label{fig5}}
\end{figure}

\begin{figure}
\caption{The same as Figure 5, except that the theoretical isochrones
are for the same metallicity ([Fe/H] = $-0.32$) and a range of ages from
5 to 18 Gyr. \label{fig6}}
\end{figure}

\begin{figure}
\caption{The luminosity function for the NGC 5128 red giant branch.
The hatched region is for the 4120 stars more metal-poor than [Fe/H]$= -0.7$,
while the clear region is for all data (10352 stars).  The adopted level
of the RGB tip at $I = 24.1 \pm 0.1$ is indicated.
 \label{fig7}}
\end{figure}
 
\begin{figure}
\caption{$I,(V-I)$ CMDs for NGC 5128 (this paper), the Local
Group dwarf elliptical NGC 147 (\protect\cite{han97}), and two dwarf
ellipticals in the M81 group (\protect\cite{cal98}).  The number
of star in each CMD is labelled in the lower right corner of eac
panel.  Fiducial lines 
for the metal-poor globular cluster M15 and the metal-rich cluster
NGC 6553, as in Fig.~4, are superimposed on each one.  NGC 147 and
M81-F8D1 have a modest mixture of low-to-intermediate metallicity,
while the very small dwarf M81-BK5N is entirely metal-poor.
\label{fig8}}
\end{figure}

\begin{figure}
\caption{The $(V-I)$ color distribution across the giant branch of
NGC 5128, in 0.5-magnitude bins from $I = 24.0 - 26.5$.  \label{fig9}}
\end{figure}

\begin{figure}
\caption{The [Fe/H] metallicity distribution in three magnitude
bins across the giant branch, obtained by interpolation within the fiducial
RGB sequences as described in the text.  Error bars in the top
panel represent the uncertainties in [Fe/H] resulting from the photometric
measurement uncertainty in the $(V-I)$ colors.  \label{fig10}}
\end{figure}

\begin{figure}
\caption{{\it Upper panel:} 
The [Fe/H] distribution for the two brightest samples in Fig.~10.
{\it Lower panel:} 
The [Fe/H] distribution for 47 globular clusters in the halo of
NGC 5128 (\protect\cite{hghh92}).  The Gaussian curves superimposed
on the metal-rich components in each graph have identical means
[Fe/H](peak) $= -0.32$ and dispersions $\sigma = 0.22$.
 \label{fig11}}
\end{figure}

\begin{figure}
\caption{The metallicity distribution function from Fig.~11, replotted
as the number of stars per unit linear abundance $Z/Z_{\odot}$.
\label{fig12}}
\end{figure}

\begin{figure}
\caption{Metallicity distribution function from Figure 11, fitted by
a two-component enrichment model as described in the text.
The low-metallicity component has an initial abundance $Z_0 = 0$
and a yield $y = 0.002 \simeq 0.12 Z_{\odot}$.  The high-metallicity
component has an initial abundance $Z_0 = 0.003$ (equivalent to 
[Fe/H]$ \simeq -0.75$) and a yield $y = 0.005$.
\label{fig13}}
\end{figure}

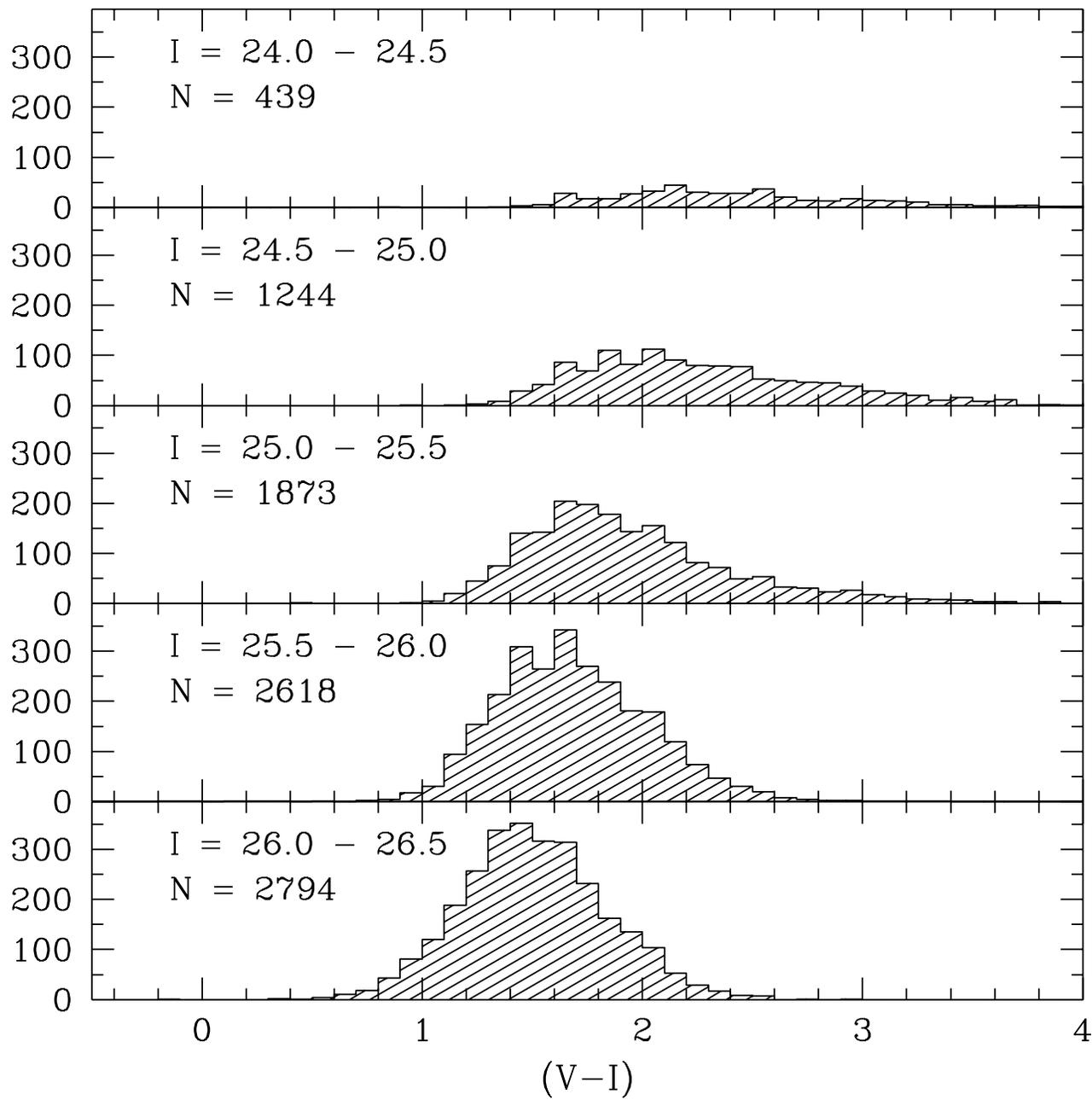
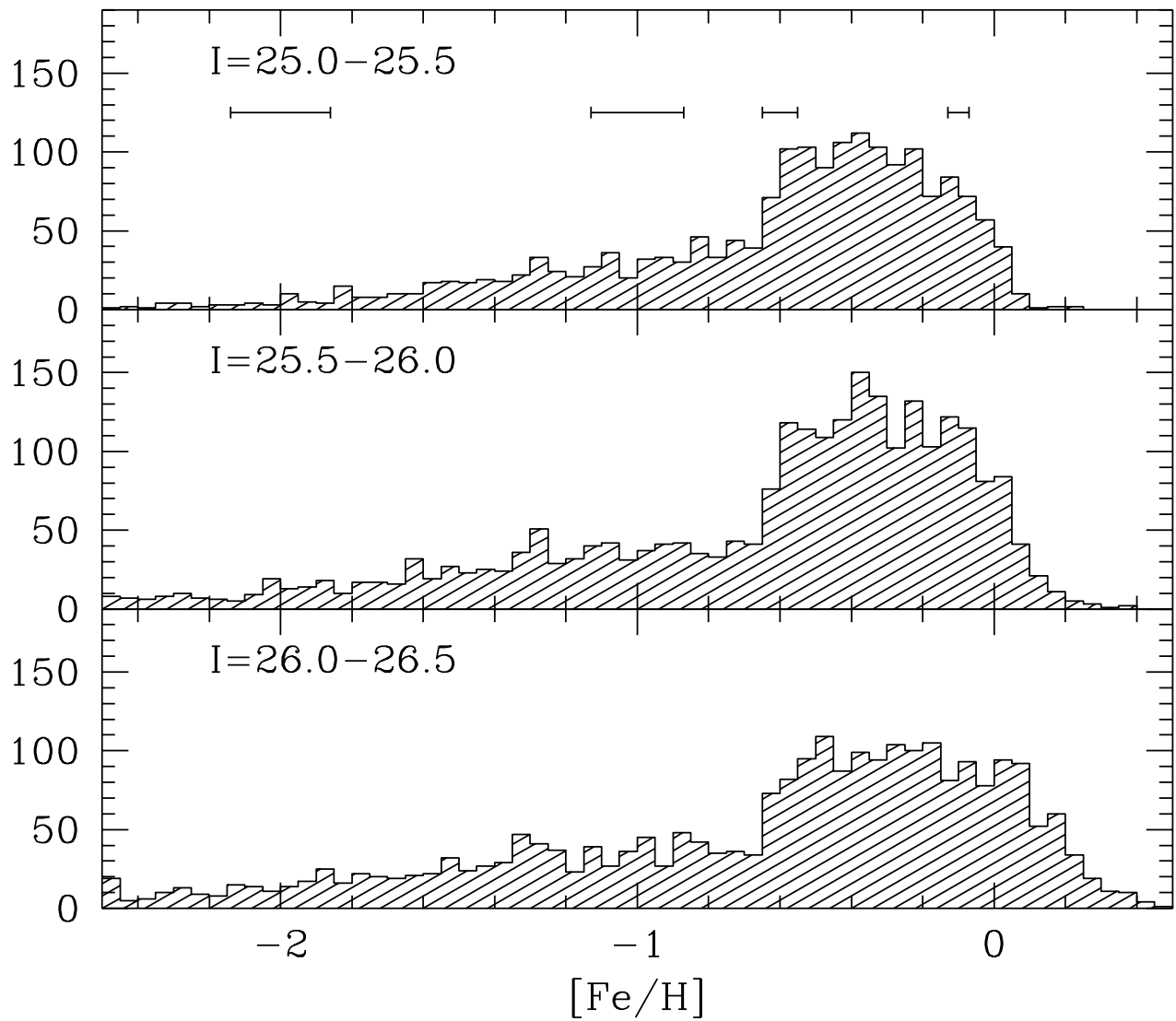
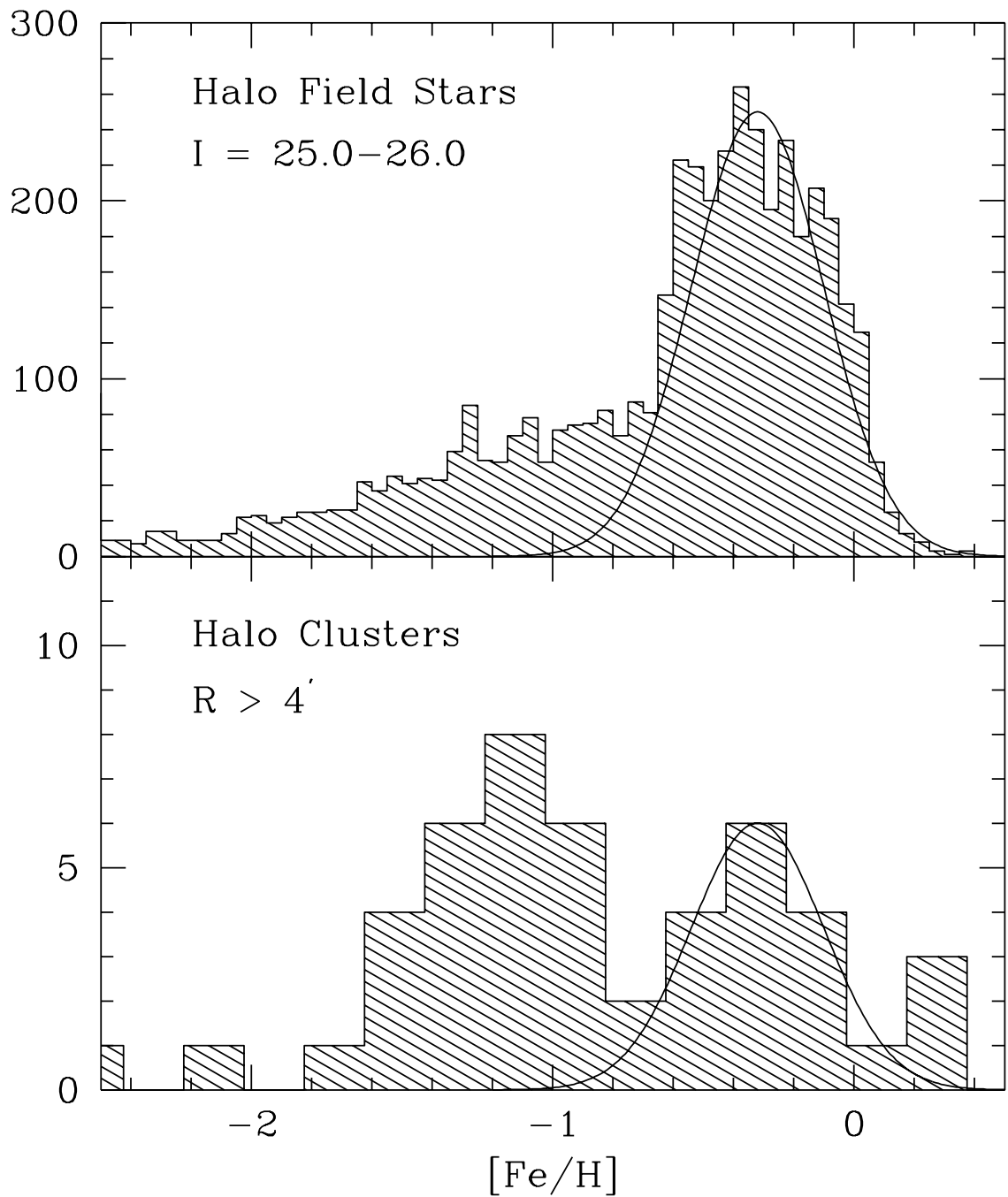
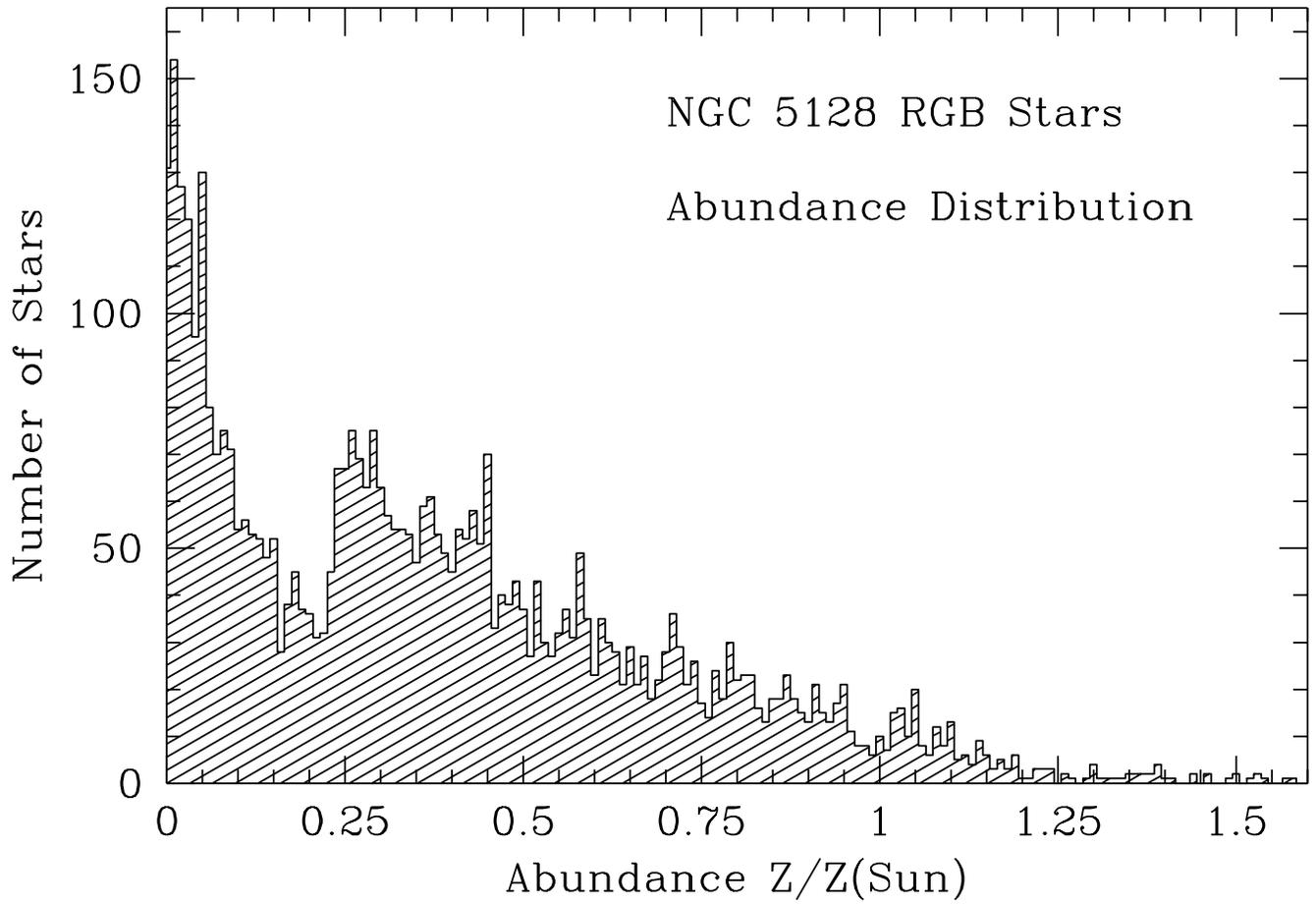
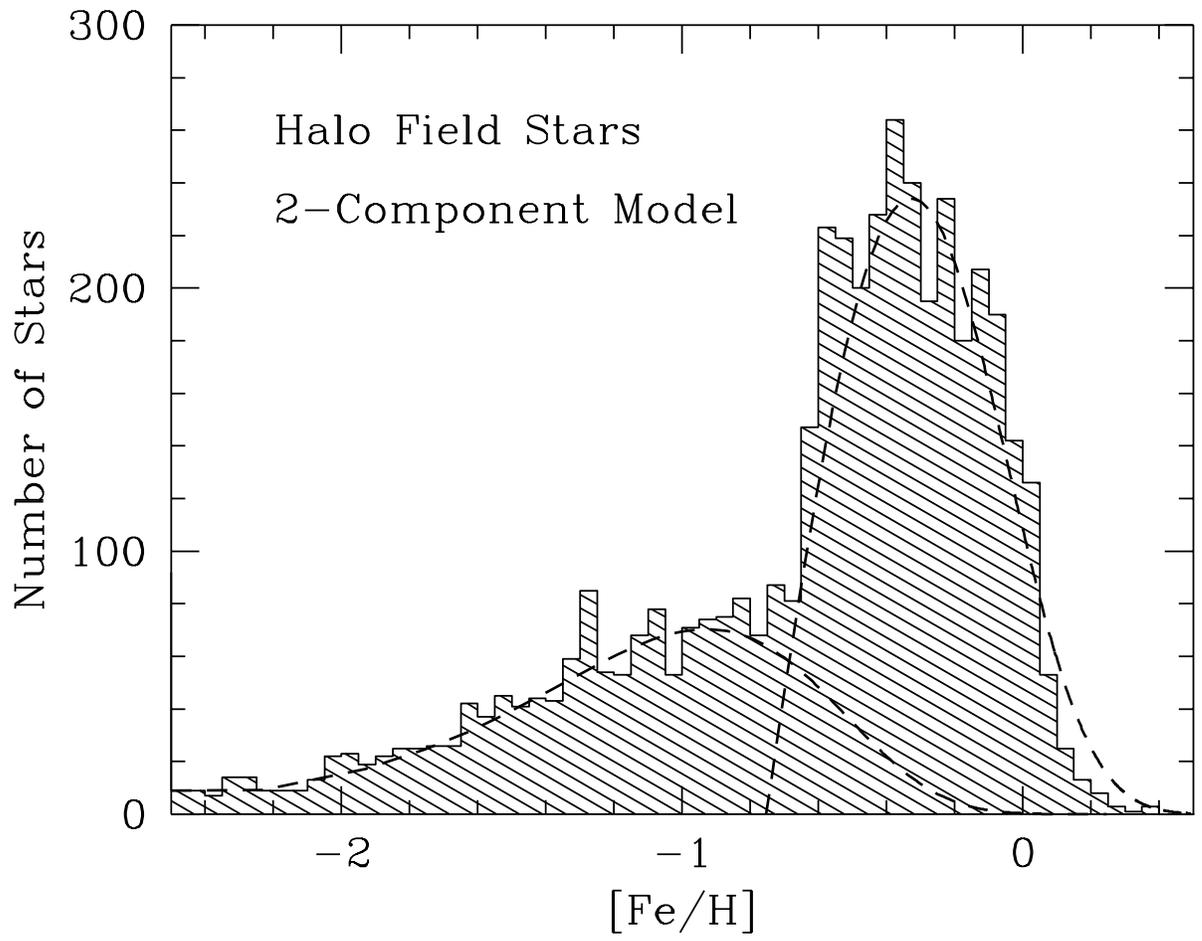
\end{document}